\documentclass[12pt]{article}
\pdfoutput=1
\usepackage[yyyymmdd,hhmmss]{datetime}
\usepackage{hyperref}
\usepackage{cite}
\usepackage{color}
\usepackage{graphicx}
\usepackage{amsmath}
\usepackage{amssymb}
\usepackage{xspace}

\makeatletter
\@addtoreset{equation}{section}

\makeatletter
\renewcommand\section{\@startsection {section}{1}{\z@}%
                                   {-3.5ex \@plus -1ex \@minus -.2ex}
                                   {2.3ex \@plus.2ex}%
                                   {\normalfont\large\bfseries}}
\renewcommand\subsection{\@startsection{subsection}{2}{\z@}%
                                     {-3.25ex\@plus -1ex \@minus -.2ex}%
                                     {1.5ex \@plus .2ex}%
                                     {\normalfont\bfseries}}

\def\baselinestretch{1.2}
\newcommand{\be}{\begin{equation}}
\newcommand{\ee}{\end{equation}}
\newcommand{\beq}{\begin{eqnarray}}
\newcommand{\eeq}{\end{eqnarray}}

\newcommand{\gone}[1]{{}}

\begin{document}
\begin{titlepage}

\rule{0ex}{0ex}

\vfil

\begin{center}

{\bf \Large
  Stability and boundedness in AdS/CFT\\ with double trace deformations \\
  II: Vector Fields
}

\vfil

William Cottrell$^1$,  Akikazu Hashimoto$^2$,\\
 Andrew Loveridge$^2$, and Duncan Pettengill$^2$

\vfil

{}$^1$ Institute for Theoretical Physics Amsterdam, University of Amsterdam\\ 1098 XH Amsterdam, The Netherlands

{}$^2$ Department of Physics, University of Wisconsin, Madison, WI 53706, USA

\vfil

\end{center}

\begin{abstract} We extend the analysis of boundedness and stability, initiated for scalar fields in anti de Sitter space in a previous work, to the case of vector fields. We show that the double trace deformation of Marolf and Ross is distinct from the double trace deformation of Witten. The former gives rise to an $SL(2,{\bf R})$ family of theories whereas the latter gives rise to an independent $SL(2,{\bf Z})$. We analyze the finite temperature two-point correlation function of current operators and infer the susceptibility and spectrum of low lying states. We discuss various physical features exhibited by these theories. 
\end{abstract}
\vspace{0.5in}

\end{titlepage}
\renewcommand{\baselinestretch}{1.05}  

\section{Introduction}

In a recent article \cite{Casper:2017gcw}, we surveyed the double
trace deformation of scalar fields in an anti de Sitter Schwarzschild
background in the probe approximation and mapped out the regions of
dynamical and thermodynamic stability in the theory space. The main conclusions reported  in \cite{Casper:2017gcw} are that
\begin{enumerate}
  \item The space of theories is naturally parameterized by the group manifold of $SL(2,{\bf R})$ which can be visualized as $AdS_3$.
  \item The full $SL(2,{\bf R})$ structure is required once one includes $S$ and $T$ deformations (corresponding to Legendre transform and the double trace deformation, respectively) as generators of deformations in the space of theories.
  \item The three generators of $SL(2,{\bf R})$ can be interpreted as parameterizing the double trace deformation, overall rescaling, and contact term deformation.
  \item Observables such as the free energy and susceptibility depend on the full set of $SL(2,{\bf R})$ parameters including the contact term, but the spectrum of small fluctuations is independent of the contact term. We referred to the positivity of the susceptibility observable as ``boundedness'' whereas the term ``stability'' referred to the absence of tachyonic fluctuations.
    \item The issue of boundedness is directly related to the observables being distributed as a normalizable distribution in the path integral.
\end{enumerate}
These conclusions can be summarized succinctly in a picture which we
illustrated in figure 2 of \cite{Casper:2017gcw}.

In this article, we examine an extension of the analysis in
\cite{Casper:2017gcw}, which was strictly in the context of scalar
fields, to a setup involving vector fields.  It was already noted in
\cite{Casper:2017gcw} that parallel issues were described in
\cite{Witten:2003ya,Marolf:2006nd} and one might think that there is
nothing left to discuss. Upon closer examination, one finds that
\cite{Witten:2003ya} and \cite{Marolf:2006nd} are discussing double
trace deformations of different kinds, giving rise to different
physics. The goal of this paper, therefore, is to spell out the
differences in the physics of \cite{Witten:2003ya} and
\cite{Marolf:2006nd} and to further compare the physics to that of the
scalar case considered in \cite{Casper:2017gcw}. It should be stressed
that the physics of Witten's double trace deformation
\cite{Witten:2003ya} is intimately connected to the $SL(2,{\bf Z})$
modular transform of Chern-Simons theory and has been studied
extensively in the context of studying the quantum Hall effect. A
sample of such work includes
\cite{Zee:1996fe,Burgess:2000kj,Leigh:2003ez,Hartnoll:2007ai,Fujita:2012fp}. We
will show that the Marolf-Ross deformation will naturally give rise to an
$SL(2,{\bf R})$ space of theories. These appear to have been studied
more in the context of generalizing boundary dynamics
\cite{Amsel:2008iz,Compere:2008us,Andrade:2011dg,YingZhao}.

The organization of this paper is as follows. In section 2, we review
the boundedness and stability issues for the scalars
\cite{Casper:2017gcw} and establish some notation. In section 3, we
formulate the double trace and contact term deformation for the vector
fields generalizing \cite{Casper:2017gcw}. We emphasize that the
prescription of Marolf and Ross \cite{Marolf:2006nd} and the
prescription of Witten \cite{Witten:2003ya} give rise to distinct
deformations, and work out the finite temperature two-point correlation
function in each of these cases. In section 4, we comment on the
physical features exhibited by these deformations and present various
concluding remarks. In appendix A, we review the basic formulation of
AdS/CFT involving vector fields.

\section{Boundedness and stability of double trace deformations for scalar fields}

In this section, we will briefly review the analysis and the
conventions for the scalars that was done in
\cite{Casper:2017gcw}. Readers are referred to \cite{Casper:2017gcw}
for a more detailed account.  We have also included a review of
standard AdS/CFT conventions in appendix \ref{AppA}.  Let us start by
recalling the action\footnote{We have however reparameterized $u =
  z_0/z$.}
\be 
S =   {1 \over 2 e^2 }    \int dz\ d^d x \  \sqrt{g} \left( {1 \over 2} g^{zz} (\partial_z \phi(x,z))^2 +{1 \over 2} g^{ii} (\partial_i \phi(x,z))^2+ {1 \over 2} {m^2 \over R^2} \phi(x,z)^2 \right) \ee
where $m^2$ is dimensionless and we assign dimension $(3-d)$ to $e^2$
so that $\phi$ has dimension one.\footnote{In \cite{Casper:2017gcw},
  we assigned dimension $(d-2)/2$ to $\phi$ and treated $e^2$ as
  dimensionless.} We will regulate the volume of ${\bf R}^{d-1} \times
S_1$ by treating it as ${\bf T}^{d-1} \times S_1$ where ${\bf
  T}^{d-1}$ has the volume of $V_{d-1}$ with all of the cycles in
${\bf T}^{d-1}$ being much larger than that of the $S_1$. We will also
concentrate on the zero momentum component of $\phi$ along ${\bf
  T}^{d-1}$ by defining
\be \tilde \phi(z) = {T \over V_{d-1}} \int d^d x \, \phi(x,z), \qquad [\tilde \phi] = 1 \ee
in terms of which  the action becomes
\be
S = {R^{d-1} V_{d-1} \over 2 e^2 T}  \int_0^{z_0}  dz\  z^{-d-1} \left( {1 \over 2} f z^2(\partial_z \tilde \phi(z))^2 + {1 \over 2} m^2 \tilde \phi(z)^2  \right) \ . \ee

It is convenient to scale out $z_0$ and work in units where $z_0$ is set to one,
\be
S = {\cal N} z_0^{2+d} \int_0^1  dz\  z^{-d-1} \left( {1 \over 2} f z^2(\partial_z \tilde \phi(z))^2 + {1 \over 2} m^2 \tilde \phi(z)^2  \right) \ , \ee
where
\be {\cal N} = {R^{d-1} V_{d-1} \over 2 e^2 T   z_0^{2+d}} \label{calN}
\ee
is dimensionless. We can now reproduce the generating function given in (3.11) of \cite{Casper:2017gcw} as follows
\beq 
Z[J] &=& \int [D \rho] [D \tilde \phi(z_c)] [D \tilde \phi]_{\tilde \phi(z_c)} \exp \left[ -{\cal N}   \int_{z_c}^1 dz\  z^{-d-1} \left( {1 \over 2} f z^2(\partial_z \tilde \phi(z))^2 + {1 \over 2} m^2 \tilde \phi(z)^2  \right)\right.
  \cr && \qquad  - {{\cal N} \over 2}  \Delta_- z_c^{-d} \tilde \phi(z_c)^2 \label{genZ4} \\
  &&  \qquad \left.\rule{0ex}{3ex}\left. 
  + {\cal N} (\Delta_+ - \Delta_-) \left( \alpha z_c^{-2 \Delta_-} \tilde \phi(z_c)^2 + z_c^{-\Delta_-}\beta (\tilde \phi(z_c)+\rho) J - {\beta^2 z_c^{-2 \Delta_-} \over 4 \gamma} \rho^2 \right)\right|_{z_{c} \rightarrow 0}  \right] \nonumber \eeq
where $z_c$ is the UV cut-off for implementing holographic renormalization \cite{Bianchi:2001kw}, and the term in the second line is the standard holographic
renormalization counter-term. Also, we have
\be \Delta_{\mp} = {d \mp \sqrt{d^2 + 4 m^2} \over 2} \ . \ee
The measure $[D\tilde \phi]_{\tilde \phi(z_c)}$ indicates integrating
over $\tilde \phi(z)$ with the boundary value at $z=z_c$ fixed in the
path integral, although we are also explicitly integrating over $\tilde
\phi(z_c)$ as well. Variation of $Z[J]$ with respect to $J$ will
encode the correlation function of operators sourced by $J$.

With $\alpha=\gamma=0$ and $\beta=1$, this generating function is
interpretable as the Legendre transform, where we integrate over the
boundary value $\tilde \phi(z_c)$, of the usual AdS/CFT prescription
in the Dirichlet setup, so it is the Neumann generating
function. Parameters $\alpha$ and $\gamma$ deform this generating
function. The parameter $\alpha$ corresponds to the double trace
deformation, while the parameter $\gamma$ controls the coupling between $J$ and
an auxiliary field $\rho$ with vanishing kinetic term.  This gives
rise to a contact-term, which can be made more manifest by integrating
out $\rho$ to write an effective generating function of the form
\beq 
Z[J] &=& \int [D \tilde \phi(z_c)]  [D \tilde \phi]_{\tilde \phi(z_c)} \exp \left[ -{\cal N}   \int_{z_c}^1 dz\  z^{-d-1} \left( {1 \over 2} f z^2(\partial_z \tilde \phi(z))^2 + {1 \over 2} m^2 \tilde \phi(z)^2  \right)\right.
  \cr && \qquad  - {{\cal N} \over 2}  \Delta_- z_c^{-d} \tilde \phi(z_c)^2 \label{genZ5} \\
  &&  \qquad \left.\rule{0ex}{3ex}\left. 
  + {\cal N} (\Delta_+ - \Delta_-) \left( \alpha z_c^{-2 \Delta_-} \tilde \phi(z_c)^2 + z_c^{-\Delta_-}\beta \tilde \phi(z_c) J +  { z_c^{2 \Delta_-} \gamma} J^2 \right)\right|_{z_{c} \rightarrow 0}  \right] \nonumber \ . \eeq

The reason we introduce the auxiliary field $\rho$ in (\ref{genZ4}) is to satisfy the requirement of the fluctuation dissipation theorem that the coupling to the source $J$ be linear, but nothing prevents us from integrating out $\rho$ for the purpose of computing the correlation functions, giving rise to a term quadratic in $J$ in the generating functional which is manifestly a contact term. 

The small $z$ asymptotics of $\tilde \phi(z)$ can be parameterized as
\be \tilde \phi(z) = p_1 z^{\Delta_-} (1 + {\cal O}(z)) - p_2 z^{\Delta_+} (1 + {\cal O}(z))\ .  \ee
In terms of these expressions, one can read off the boundary condition by varying with respect to $\tilde \phi(z_c)$, and the degree of freedom sourced by $J$ by varying with respect to $J$.  This can be expressed in the form
\beq
      {1  \over (\Delta_+ - \Delta_-) {\cal N}} {\delta \over \delta J}  & = & \left(\beta  
    -{4 \alpha  \gamma \over \beta}\right) p_1 + {2\gamma \over \beta} p_2 \label{ddj} \\
     J &=& -{ 2  \alpha  \over \beta} p_1 + {1 \over \beta} p_2  \label{boundaryc} \ , 
\eeq
which simplifies to the expected Neumann expression for $\alpha=\gamma=0$ and $\beta=1$.

One important feature  implicit in (\ref{ddj}) and (\ref{boundaryc}) is the fact that the expressions in terms of $\alpha$, $\beta$, and $\gamma$ appearing on the right hand side have a natural $SL(2,{\bf R})$ parameterization under the map
\be
\Lambda=\left(\begin{array}{cc} a & b \\ c & d\end{array}\right) =
\left(\begin{array}{cc}        \beta  
     -{ 4 \alpha  \gamma \over \beta}&  {2\gamma \over \beta} \\
 -{2 \alpha  \over \beta}  & {1 \over \beta}\end{array}\right) \ . \ee

The factor of $ (\Delta_+ -\Delta_-)$ is a bit annoying, but note that
this is a {\it positive} number of order one.  Taking slight liberty
in the normalization conventions, we can define the ``operator''
\be {\cal O} \equiv   {1\over (\Delta_+ - \Delta_-)   {\cal N}} {\delta \over \delta J} \ee
and infer that the normalized correlation function is
\be \chi = (\Delta_+ - \Delta_-) {\cal N} \langle {\cal O}{\cal O} \rangle  = {\delta {\cal O} \over \delta J} = {a p_1 + b p_2  \over c p_1 + d p_2} \ . \ee
We refer to this quantity as susceptibility. 
For the Dirichlet and Neumann theories, this correlation function simplifies to
\be \chi_N = {p_1 \over p_2}, \qquad \chi_D = -{p_2 \over p_1} \ee
and so we can write the susceptibility for the general case in the form
\be \chi_\Lambda = {a \chi_N + b \over c \chi_N + d} = {a - b \chi_D \over c  - d \chi_D} \ . \ee
In other  words, the susceptibility transforms modularly.

Note that despite formally being an expectation value of the square of an
operator, the susceptibility $\chi_\Lambda$ can be positive or negative. In
fact, the fact that $\chi_N$ and $\chi_D$ are related by an $S$
transformation of $SL(2,{\bf R})$ requires that one be positive and
the other be negative. The point of \cite{Casper:2017gcw} was to shed
light on this issue.

It is straightforward to extend this analysis to the case when momentum along
${\bf T}^{d-1}$ is non-vanishing. This is  explained in detail in
section 3 of \cite{Casper:2017gcw}. One can also infer the spectrum of
small fluctuations by examining the poles of the two-point
function. For finite $T$, the spectrum is gapped as is expected. For
some set of boundary conditions in the $SL(2,{\bf R})$ parameter
space, the spectrum will include a normalizable tachyon. These
features are summarized in figure 2 of \cite{Casper:2017gcw}.

\section{Boundedness and stability of double trace deformation for vector fields}

We are now ready to consider the generalization of
\cite{Casper:2017gcw} to vector fields.  A useful and concrete
starting point is to write the analogue of (\ref{genZ5}). Already, at
this point, the differences between the construction of Witten
\cite{Witten:2003ya} and Marolf and Ross \cite{Marolf:2006nd} emerge.

\subsection{Marolf-Ross v.s.\ Witten}

Following the template of  \cite{Casper:2017gcw}, consider the generating function
\be Z[K_i] = \int [D a_i] [D A_i]_{a_i} \exp\left[ - S[A_i] + S_{CT}[a_i]+ \int d^dx\ (d-2) (\alpha a_i a_i + \beta a_i K_i + \gamma  K_i K_i)  \right] \label{defineabc} \ee
where $S[A_i]$ is the positive definite Maxwell action (\ref{action})
and $S_{CT}$ is the holographic renormalization counter-term. We
could, if desired, integrate back in an auxiliary vector field $\rho_i$
to make the coupling to the source $K_i$ linear. The coupling constant
$e^2$ in (\ref{action}) has dimensions
\be [e^2] = 3-d \ee
so that the gauge field $A_i$ and the source $K_i$ have dimensions
\be [A_i] = 1 , \qquad [K_i] = d-1 \ee
and
\be [\alpha]=d-2, \qquad [\beta]=0, \qquad [\gamma] = 2-d \ . \ee

Aside from the contact term parameterized by $\gamma$, this is essentially the type of generating function considered in  \cite{Marolf:2006nd}.

The formulation of Witten's double trace deformation \cite{Witten:2003ya} is different. It is a formulation which only exists when $d=3$, and can be written in the form
\be Z[K_i] = \int [D a_i]  [D A_i]_{a_i} \exp\left[ - S[A_i] + i \int d^3 x \, (\alpha \epsilon^{ijk} a_i \partial_j a_k + \beta \epsilon^{ijk} a_i \partial_j K_k  + {\gamma} \epsilon^{ijk} K_i \partial_j K_k )  \right]  \ . \label{witten} \ee
This time, the dimensions are such that
\be [\alpha] = [\beta] = [\gamma]=0, \qquad [A] = [a] = [K] = 1\ . \ee
Although (\ref{defineabc}) and (\ref{witten}) are similar in
structure, and seem completely natural as the generalization of
(\ref{genZ5}), they are manifestly different in the details. The
goal of this section is to explain the features such as stability,
boundedness, and the role of contact terms, for both of these
scenarios.

\subsection{Double trace deformation of Marolf and Ross}

Let us now take a closer look at the structure of the generating function
(\ref{defineabc}). When $\alpha=\gamma=0$ and $\beta=1$, this is
interpretable as the Legendre transform from the standard Dirichlet
formulation to the Neumann one. One can think of this as gauging the
$U(1)$ global symmetry of the original CFT and coupling the $U(1)$
gauge field $a_i$ to the source $K_i$.  To the extent that the
Legendre transform involves a path integral over a vector field $a_i(x)$
on the boundary, one must formally prescribe the quotient of the gauge
orbit.  On the first pass, this may seem problematic in that the
double trace term $\alpha a_i a_i$ is manifestly not gauge invariant.
This can be remedied by thinking of the double trace term as arising
from a Stueckelberg action \cite{Ruegg:2003ps}
\be \alpha a_i a_i \rightarrow \alpha (\partial_i \varphi - a_i)(\partial_i \varphi - a_i) \label{stueckelberg}  \ee
and including the path integral over the Stueckelberg field $\varphi$. Then, one can formally define the quotient by gauge orbit using the standard Fadeev-Poppov procedure. For the purpose of working at the classical level, we can also just as well chose the gauge where $\varphi = 0$.

From the expression (\ref{defineabc}) for the generating function, we can infer the operator and expectation value relation
\beq
{1 \over {\cal N} (d-2)} {\delta \over \delta \tilde K_i}& = &{ (\beta^2 - 4 \alpha \gamma)  \over \beta} \tilde a_i 
- {2 \gamma \over \beta}  {1 \over (d-2)} \left({1 \over   e^2} \sqrt{g} g^{zz} g^{ii} \partial_z \tilde A_i\right)  \label{partialK} \\
 \tilde K_i &=&   - {2  \alpha  \over \beta}  \tilde a_i -{1\over \beta} {1 \over  (d-2) }  \left({1 \over e^2}  \sqrt{g} g^{zz} g^{ii} \partial_z \tilde A_i\right)\ . \label{theK} 
\eeq
We recognize the same $SL(2,{\bf R})$ structure we saw for the scalars by identifying
\be \Lambda= \left(\begin{array}{cc} a & b \\ c & d \end{array}\right) =
\left(\begin{array}{cc} { (\beta^2 - 4 \alpha \gamma)  \over \beta} &
{2 \gamma \over \beta}   \\
- {2  \alpha  \over \beta}  &  {1\over \beta} \end{array}\right) \ . 
\ee

Upon identifying
\be p_{1i} =  \tilde a_i, \qquad  p_{2i} = - {1 \over (d-2)}  {1 \over e^2} \sqrt{g} g^{zz} g^{ii} \partial_z \tilde A_i \ee
we see that
\be \tilde A_i =  p_{1i} - p_{2i} z^{d-2} + \ldots \label{Aexpand} \ee
and that
\be \Delta_{ij} = {a p_{1i} + b p_{2i} \over c p_{1j} + d p_{2j}}
= \left(a \delta_{ik}  - b \Delta^D_{ik} \right) \left( c \delta_{jk} - d \Delta^D_{jk}\right)^{-1}
\ee
where
\be \Delta^D_{ik}= - {\partial p_{2i} \over \partial p_{1j}} \ . \ee
The factor of $(d-2)$ is the analogue of $(\Delta_+ - \Delta_-)$ which we encountered in the scalar case.  We see that the case $d=2$ is special because the expansion (\ref{Aexpand}) degenerates. This is an indication of strongly coupled physics in the IR and can be addressed by including logarithmic terms and a holographic renormalization counterterm \cite{Hung:2009qk,Cottrell:2015kra}. 

The conclusion of this analysis is that $\Delta_{ij}$ transforms modularly as a matrix. Focusing on the components enumerated in (\ref{2x2}) and using the fact that $J=H=0$ for the Dirichlet pure Maxwell theory, we find that the $SL(2,R)$ orbit has $F_\Lambda$ and $G_\Lambda$ of the form
\be F_\Lambda = {a - b F_D \over c - d F_D}, \qquad G_\Lambda = {a - b G_D \over c - d G_D} \ . \label{FGModular} \ee 
In other words, $F_\Lambda$ and $G_\Lambda$ transform separately and modularly.  We can now use the result of the computation of $F_D$ and $G_D$ in appendix \ref{AppA} and infer the susceptibility and the spectrum of normalizable states for each of the theories in the space of theories parameterized by  $SL(2,{\bf R})$.

\begin{figure}
\centerline{  \begin{tabular}{cc}\includegraphics[width=2in]{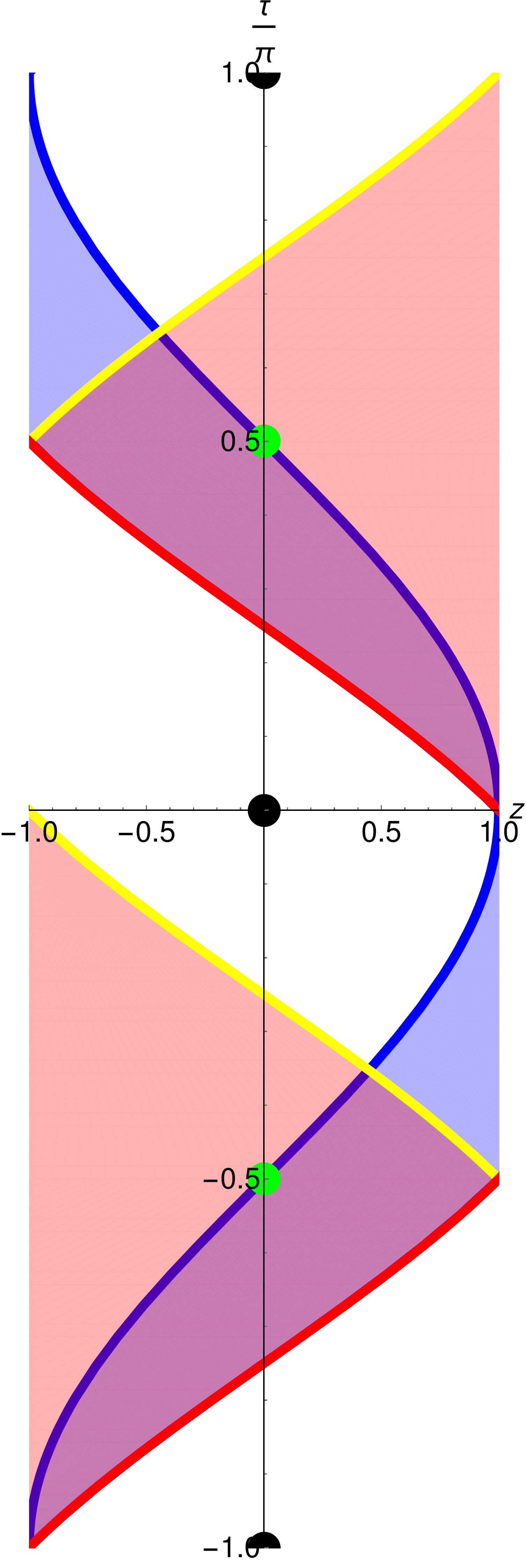} &\includegraphics[width=2in]{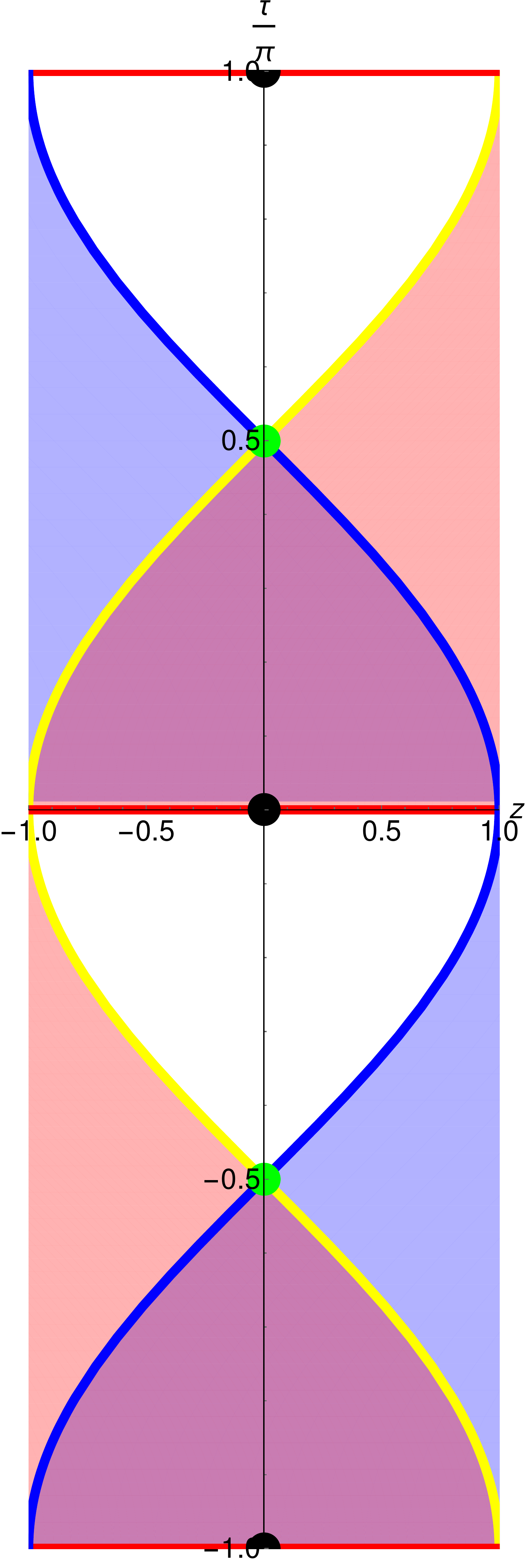}\\
      (a) & (b) \end{tabular}}
\caption{Cross section plot of $SL(2,{\bf R})$ theory space for the Marolf-Ross family of theories. Plot (a) encodes the longitudinal component and plot (b) encodes the transverse component for $\omega=0$. The black dot corresponds to the Neumann point and the green dot corresponds to the Dirichlet point. Inside the red shaded region, the susceptibility is negative and we say that the model is unbounded. Inside the blue region, the spectrum of small fluctuation includes a tachyon. The tachyons for the longitudinal component of the vector fluctuations in ${\bf R}^{d-1}$ are ghostlike when continued to Minkowski signature ${\bf R}^{1,{d-2}} \times S_1$. A similar plot for the scalar $SL(2,{\bf R})$ family of scalar theories can be found in figure 2 of \cite{Casper:2017gcw}.
  \label{fig:MR}}
  \end{figure}

We can summarize the essential features of the $SL(2,{\bf R})$ family of
models arising from the Marolf-Ross deformations by producing a plot
similar to figure 2 of \cite{Casper:2017gcw}. This is provided in
figure \ref{fig:MR}. To avoid cluttering the image, we have only
illustrated the cross-section, parameterized by $\tau$ and $z$, of the
$SL(2,{\bf R})$ theory space along the plane where $\phi=0$ and
$\phi=\pi$ where $\tau$, $\phi$, and $\rho = \tanh^{-1} z$ are the standard cylindrical coordinates on  $AdS_3$. Figure \ref{fig:MR}.a
illustrates the features of the longitudinal components  encoded in
$F_\Lambda(p^2,\omega)$, for $\omega=0$. Just as in
\cite{Casper:2017gcw}, the red shaded region is where the
susceptibility $\chi = F_\Lambda(0)$ is negative and consequently,
the correlation function is unbounded. The shaded blue region is where the
spectrum contains a tachyon. Along the red line, where the shaded blue
region and the shaded red region share a boundary, one expects a
normalizable massless fluctuation. Figure \ref{fig:MR} is in fact
identical to what was drawn in figure 2 of
\cite{Casper:2017gcw}. Figure \ref{fig:MR}.b, on the other hand,
illustrates the same features but for the transverse components
encoded in $G_\Lambda(p^2,\omega)$, again for $\omega=0$. This time,
the red line intersects the Neumann point. This is where one might
expect to find a massless vector in the spectrum. We will comment on
the potential subtleties involved with approaching the massless point as well as
working in the tachyonic region when we discuss the physics of these
models in the following section.

\subsection{Double trace deformation of Witten}

Let us now examine the behavior of generating function (\ref{witten}) for the double trace deformation of Witten type \cite{Witten:2003ya}. This is a formulation which only exists in $d=3$ because that is the only dimension in which an anti-symmetric 3 form exists as an invariant tensor. The action (\ref{witten}) should be viewed as an effective action, obtained by considering a general theory of the type
\be i \int d^3 x \left(
    {n_1 \over 4 \pi} \epsilon^{i j k} A_i \partial_j A_k +
    {1 \over 2 \pi} \epsilon^{i j k} A_i \partial_j B_k +
    {n_2 \over 4 \pi} \epsilon^{i j k} B_i \partial_j B_k +
    {1 \over 2 \pi} \epsilon^{i j k} B_i \partial_j K_k \right) \label{micro}
\ee
where $n_1$ and $n_2$ are integers, as is required by the quantization of Chern-Simons level. One can also consider repeating this manipulation for any finite set sequence of integers $(n_1, n_2, \ldots n_N)$.  Integrating out all fields except $A$ (and recalling that $K$ is not a field), one finds that
\be \Lambda= \left(\begin{array}{cc} a & b \\ c & d \end{array}\right) =
\left(\begin{array}{cc} { 2 \pi (\beta^2 - 4 \alpha \gamma)  \over \beta} &
{2 \gamma \over \beta}   \\
- {2  \alpha  \over \beta}  &  {1\over 2 \pi \beta} \end{array}\right)
\ee
must be an element of $SL(2,{\bf Z})$. This implies that $\alpha$,
$\beta$, and $\gamma$ can take on fractional values. This is fine as
long as one understands that this is an effective theory. See
\cite{Bergman:1999na,Gaiotto:2008ak,Closset:2012vg,Closset:2012vp} for
related discussions. One should also regard (\ref{micro}) as the
microscopically well defined prescription for introducing the
auxiliary field (which we refer generally as $\rho$) in order for the
source $K$ to couple linearly to the fields in order to satisfy the
assumptions of the fluctuation dissipation theorem.

In order to understand the $SL(2,{\bf Z})$ transformation of the susceptibility and the two-point function, we need the analogue of (\ref{partialK}) and (\ref{theK}).  It is not very difficult to show that (\ref{partialK}) and (\ref{theK}) for (\ref{witten}) becomes
\beq
{1 \over {\cal N} (d-2)} {\delta \over \delta \tilde K_i}& = &{ (\beta^2 - 4 \alpha \gamma)  \over \beta} \epsilon_{ijk} p_j \tilde a_k 
- {2 \gamma \over \beta}  {1 \over (d-2)} \left({1 \over   e^2} \sqrt{g} g^{zz} g^{ii} \partial_z \tilde A_i\right)  \label{WittendK} \\
\epsilon_{ijk} p_j \tilde K_k &=&   - {2  \alpha  \over \beta}  \epsilon_{ijk} p_j\tilde a_k -{1\over \beta} {1 \over  (d-2) }  \left({1 \over e^2}  \sqrt{g} g^{zz} g^{ii} \partial_z \tilde A_i\right)\ . \label{WittenK} 
\eeq
With a little bit of algebra, one can show that the quantity
\be \sigma_{ij} = P_{mn} p_k \epsilon_{kjm} \Delta_{in} \ee
transforms modularly, in terms of which
\be \Delta_{ij} = - \sigma_{il} \epsilon_{lmj} p_m \ . \ee
The quantity $2 \pi \sigma_{ij}$ indeed can be interpreted as the
conductivity tensor which is known to transform modularly
\cite{Zee:1996fe,Burgess:2000kj,Hartnoll:2007ai,Fujita:2012fp}. One
can therefore start with $\Delta_{ij}$ for the Dirichlet theory, map
to $2 \pi \sigma_{ij}$, perform the modular transformation, and map
back to $\Delta_{ij}$. This time, the $SL(2,{\bf Z})$ transform can
give rise to non-vanishing $H$ in (\ref{2x2}) even if $H=0$ for the
Dirichlet theory. Curiously, the symmetric component $J$ appears to
remain zero for these classes of models.

One can in fact write $\sigma$ for the components of (\ref{2x2}) in the form
\be \sigma = 
{1\over p} \left[\begin{array}{cc}  pH & -F \\ G &  pH \end{array} \right] \ . 
\ee
Note that in the zero temperature limit, $F$ and $G$ become the same, and $\sigma$ can be viewed as
\be  H {\bf 1} + {1 \over p}   F (i\sigma_2) \leftrightarrow {1 \over 2 \pi } (w + it) \ee
in the notation of \cite{Witten:2003ya}. It is $w + it$ which transformed modularly in \cite{Witten:2003ya}. We can understand our computation as simply the finite temperature generalization of the computation of \cite{Witten:2003ya}.

It is straightforward to read off the form of $F$, $G$, and $H$ for any of the models labeled by elements of $SL(2,{\bf Z})$. We find that
\beq
F_\Lambda(p^2) &=& {p^2 F_D(p^2)  \over c^2 p^2 + 4 \pi^2 d^2 F_D(p^2) G_D(p^2)} \label{Fwitten} \\
G_\Lambda(p^2) &=& {p^2 G_D(p^2)  \over c^2 p^2 + 4 \pi^2 d^2 F_D(p^2) G_D(p^2)} \label{Gwitten}\\
H_\Lambda(p^2) &=& {a c p^2 + 4 \pi^2 b d F_D(p^2) G_D(p^2) \over 2 \pi c^2 p^2 + 8 \pi^3 d^2 F_D(p^2) G_D(p^2)} \ . \label{Hwitten}
\eeq
Note that although strictly speaking $\{a, b, c, d\}$ should form a discrete set consistent with the $SL(2,{\bf Z})$ structure, the formulas for $F_\Lambda(p^2)$, $G_\Lambda(p^2)$, and $H_\Lambda(p^2)$ depend smoothly on them, as if they were defined on $SL(2,{\bf R})$. This is not uncommon for a physical observable in fractionally structured systems \cite{Hofstadter:1976zz}.

One might consider drawing a diagram similar to figure \ref{fig:MR} for the Witten theory, but it appears that in this class of theories we never encounter models containing tachyons. Curiously, the susceptibility
\be \chi_\Lambda =   -{1 \over c^2 + 4 \pi^2 d^2} \ee
is strictly negative.  We will comment on what these things might
possibly mean physically in the next section.

\section{Physical features of Marolf-Ross and Witten deformations}

Now that we have spelled out the technical aspects of Marolf-Ross and Witten deformations of vector fields in anti de Sitter space, let us comment on their physical features.

First and foremost, we wish to reiterate that the two deformations are
distinct from one another. That of Marolf and Ross gives rise to an
$SL(2,{\bf R})$ set of allowed deformed theories, whereas Witten's
gives rise to a discrete set corresponding to $SL(2,{\bf Z})$. The way
in which the modular transform acts is also different. The two deformations give rise to
different boundary conditions, and they ultimately give rise to
different correlation functions. Clearly, they are physically
distinct.

Following our own treatment of the scalars in \cite{Casper:2017gcw},
we formulated our analysis in Eucledian signature where the boundary
is topologically ${\bf R}^{d-1} \times S_1$ (or ${\bf T}^{d-1} \times
S_1$ for the purpose of providing an IR cut-off).  It is worth noting
at this point that in contemplating real time physics, one can either
consider Wick rotating along one of the ${\bf R}^{d-1}$ directions or
the $S_1$ direction. We will make some comments on both.

Since the $S_1$ is compact and considered to be small, it is natural
to restrict to the zero mode sector in this direction and set
$\omega=0$ for many issues.  If so, $A_0$ essentially becomes a scalar
in ${\bf R}^{d-1}$. The susceptibility, computed for the Dirichlet theory, turns out to be negative. This is in agreement with what we found for scalars in our previous work  \cite{Casper:2017gcw}.

For the case of scalars, the $SL(2,{\bf R})$ transform was found to
potentially push the susceptibility into positive values. It is natural
to contemplate what the status of this issue is for the $A_0$
field. Here, there is already a difference between Marolf-Ross and
Witten. In the Marolf-Ross case, the susceptibility transforms modularly as
shown in (\ref{FGModular}). As such, one expects to find a class of
theories in the $SL(2,{\bf R})$ theory space with positive
susceptibility. In fact, the Neumann theory has positive susceptibility.

The situation is quite different in the Witten case. Looking at
(\ref{Fwitten}) combined with the fact that $F_D(p^2)$ and
$G_D(p^2)/p^2$ at $p^2=0$ are finite and negative implies that
$F_\Lambda(0)$ is negative definite for all of the models
parameterized by $SL(2,{\bf Z})$.

It is interesting to compare this result to the charge susceptibility
of Reissner-Nordstrom black holes in anti de Sitter space. In
canonical ensemble at fixed charge, one defines the free energy
\be F(Q) \ee
which in the presence of background chemical potential is
\be F(Q) - \mu Q \ . \ee
The system seeks to adjust $Q$ to minimize this quantity, and so the stability is encoded in the condition that 
\be {d^2 F(Q) \over dQ^2} = {d \mu \over d Q} > 0\ee
which is equivalent to
\be {d Q \over d \mu} > 0 \ee
in grand canonical ensemble. The fact that large charged black holes
are stable (see e.g.\ \cite{Chamblin:1999hg}) appears to be at odds
with our observation that the susceptibility of $A_0$ is
negative. This apparent conflict can be resolved upon realizing that
in order to recover the black hole physics, we analytically continue
$A_0$ and therefore $Q$. If $F(Q) \sim -Q_E^2$ but $Q_E = i Q$,
then $F(Q) \sim Q^2$ with a positive coefficient .

The susceptibility for the transverse mode encoded in $G_D(p^2)$
appears to be automatically zero for the Dirichlet theory. In the
Marolf-Ross setup, since the susceptibility $G_\Lambda$ also
transforms modularly according to (\ref{FGModular}), we will find a
region in $SL(2,{\bf R})$ theory space where the $G_\Lambda(0)$ is
positive and a complementary region where $G_\Lambda(0)$ is
negative. The regions of positive and negative susceptibilities were
illustrated in figure \ref{fig:MR}.  In the Witten setup, one sees
according to (\ref{Gwitten}) that $G_\Lambda(0)$ is zero for all the
theories in the $SL(2,{\bf Z})$ theory space. We are beginning to see
the trend that while the behavior of the theories is sensitive to the
parameters in the Marolf-Ross setup, the Witten setup in contrast is much
more rigid.

Another class of physical features explored for the scalars in
\cite{Casper:2017gcw} is the spectrum of small oscillations. The most
straightforward case is to restrict to the $\omega=0$ sector and
identify the poles in the correlation function as a function of $p^2$
as corresponding to normalizable fluctuation modes.  In the case of
scalars, there was indeed a set of gapped normalizable states as long as
the radius of $S_1$ was kept finite. It was also observed that
tachyons can appear in the spectrum as the boundary condition is
varied. The precise region in the $SL(2,{\bf R})$ parameter space where
the spectrum includes a tachyon was illustrated in figure 2 of
\cite{Casper:2017gcw}. An especially interesting feature we can infer
from the analysis of small fluctuation of this type is the fact that
one can find massless normalizable fluctuations precisely at the
boundary of the tachyonic region. When embedded into a non-linear
setup, the appearance of a normalizable massless mode is precisely
identifiable as the field responsible for the long range correlation
at criticality. Near the onset of such a criticality, the tachyon
is a signal of unstable vacua giving rise to second order phase
transitions.

It is natural to contemplate what the status of analogous issues is
for the vectors. If one contemplates the $A_0$ component of the gauge
field from the ${\bf R}^{d-1}$ point of view, the Marolf-Ross version
of the story is very similar to what we found for the scalars. Perhaps this
should not come as a surprise by now. $A_0$ is a Kaluza-Klein scalar
from the ${\bf R}^{d-1} \times S_1$ point of view.

The situation with regards to the vectors in ${\bf R}^{d-1}$ is more
interesting. For generic $SL(2,{\bf R})$ elements, the spectrum is
gapped, as one would expect from the presence of thermal
$S_1$ \cite{Witten:1998zw}.  The lightest state can have positive or
negative mass squared. It is easy to interpret the positive mass
squared perturbations as corresponding to physical, massive, vector
fluctuations when ${\bf R}^{d-1}$ is continued to Minkowski signature
${\bf R}^{1,d-2}$.

As we explore the $SL(2,{\bf R})$ theory space and modify the boundary
condition for the bulk gauge fields accordingly, the mass of the
lowest mass state can be pushed to zero, and beyond. In fact, in the
Marolf-Ross setup, the Neumann boundary condition precisely
corresponds to the case where the vectors on ${\bf R}^{d-1}$ is
massless. This is also consistent with the observation of
\cite{Marolf:2006nd} that massless normalizable modes were found for
vectors in anti de Sitter space in global coordinates with Neumann
boundary conditions. In both the global anti de Sitter space and in
thermally compactified Poincar\'{e} AdS geometry, there is an explicit IR
cut-off and the bulk modes are expected to normalizable
\cite{Witten:1998zw}. The Neumann boundary condition appears to
precisely support such a normalizable mode.

There is, however, a subtlety with continuing the squared mass of
vector fields to zero and and to negative values in the Lorentzian
setup where the boundary is ${\bf R}^{1,d-2}\times S_1$.  The issue
concerns the longitudinal mode, which is a perfectly normalizable,
gapped state when the mass squared is positive. In the massless limit,
however, this state becomes null, and when mass squared is negative,
the state has negative norm. This issue is easier to understand in the
context of the equivalent Goldstone description
\cite{ArkaniHamed:2002sp,Peskin:2017emn}, where one uses the
Stueckelberg field instead of the longitudinal photon to
parameterize the degree of freedom. The $m^2\rightarrow 0$ limit then
makes the Stueckelberg action (\ref{stueckelberg}) strongly coupled,
and negative $m^2$ makes the action negative. For Abelian theories,
the strict $m^2 \rightarrow 0$ limit may be safe, but continuation to
negative $m^2$ looks to be problematic. With non-linearities included,
one expects even more serious singularities in taking the massless
limit, at least in flat space
\cite{vanDam:1970vg,Zakharov:1970cc}. This pathology in taking
massless limit of a vector field\footnote{A slightly different story
  \cite{Porrati:2000cp,Kogan:2000uy,Karch:2001jb} emerges when the
  cosmological constant of the boundary is non-zero.}  appears to be
consistent with the difficulty in generalizing the standard
Landau-Ginzburg description of second order phase transition as
condensation of scalars to the case of vectors. In fact, attempts to
stabilize the tachyons via quartic interactions
\cite{Hashimoto:2008tw} appear to lead to an irrelevant operator, the sign of which signals tension with
unitarity and analyticity \cite{Adams:2006sv}.  This also sits well
with the empirical observation that symmetry breaking
via vector fields acquiring a vacuum expectation value has
not yet been seen in particle physics or condensed matter contexts,
when other features such as superconductivity and the quantum Hall effect
are prevalent. We suspect similar issues are present when attempting
to induce normalizable massless spin 2 modes by tuning the double
trace deformations
\cite{Leigh:2003ez,Leigh:2007wf,Compere:2008us,Andrade:2015gja,McGough:2016lol}. Another
interesting issue to explore is whether the dichotomy between
$SL(2,{\bf R})$ and $SL(2,{\bf Z})$ exists also in the gravity
sector. We believe that the formulation (\ref{defineabc}) we adopt
 in classifying the boundary condition, which explicitly accounts
for the double trace and contact term deformations, is useful for
studying these issues systematically.

Another useful application of the parameterization (\ref{defineabc})
is to explicitly implement a Thirring $J^2$ type deformation in the
Dirichlet boundary condition.  A path integral expression for a
generating function for such a setup where one Legendre transforms
from Dirichlet, to Neumann, and back to Dirichlet again, can be
written explicitly as follows
\beq Z[\mu_i]
&=& \int [D a_i]  [D A_i]_{a_i}  [D J_i] \exp\left[ - S[A_i] + (d-2) (\beta a_i J_i +  \gamma J_i J_i- \beta J_i \mu_i )  \right] \cr
&=& \int [D a_i]  [D A_i]_{a_i}  \exp\left[ - S[A_i]   -{(d-2)\beta^2 \over 4 \gamma} (a_i - \mu_i)^2   \right] \ . 
\eeq
Here, $\mu_i$ is playing the role of the Dirichlet source. One can think of this model as an element $\Lambda$ in the $SL(2,{\bf R})$ family of theories where
\be \Lambda = \left(\begin{array}{cc} 0 & -1 \\ 1 & {2 \gamma \over \beta^2}\end{array}\right) \ . \ee
As such, this is precisely in the class of double trace deformed
Dirichlet theories discussed in \cite{Casper:2017gcw}. Note that the
act of adding a $J^2$ term can be seen as effectively inducing a
Gaussian width for the path integral of $a_i$. This deformation does
induce a normalizable, propagating, spin 1 degree of freedom although
for a generic value of $\gamma/\beta^2$, the state is massive.
Perhaps this way of thinking is closely related to how a $T \bar T$
deformation induces a normalizable gravitational fluctuation in the
analysis of \cite{McGough:2016lol}. It would be interesting if this
type of Gaussian deformation could be related to an explicit Dirichlet
wall along the lines considered in \cite{Andrade:2015gja,McGough:2016lol}. It is
also worth noting that the limit $\gamma \rightarrow 0$ may also be
subtle since it involves integrating out a vector field $J_i$ in
the process.

Not surprisingly, the Witten version (\ref{witten}) of the story for
the spectrum of small fluctuations (aside from the well established
application to quantum Hall effect) is much less exciting. Massless and
tachyonic modes do not appear to arise. Perhaps one can understand
this simply as a reflection of the fact that Chern-Simons terms are a
reliable mechanism for generating a mass gap for vector fields in
2+1 dimensions.

Let us close this paper by considering few possible interesting
extensions of this work. A main ingredient in the construction of the
models with double trace and contact term deformations of the type
(\ref{defineabc}) and (\ref{witten}) is the gauging of a $U(1)$ global
symmetry of a CFT which admits a AdS/CFT type dual. It might be
interesting to consider a setup where several CFTs with $U(1)$ global
symmetry are gauged together so that they interact via the gauge
sector. This will give rise to a holographic dual which has several
distinct IR branches which communicate via the degrees of freedom
localized in the UV. The various IR sectors are approximately
decoupled in the IR limit, and that corresponds to different branches
of the bulk solution. Presumably, upon increasing the strength of the
gauge interaction, the different branches of the gravity dual will
gradually merge, starting from the UV end. It would be interesting to
make this idea more concrete.\footnote{Of course, there are other ways to couple the AdS systems at the boundary that one could consider.}  Constructions similar to this have also
been considered in the context of Randall-Sundrum model building
\cite{Karch:2000ct}.

\begin{figure}
\center{\includegraphics[width=5in]{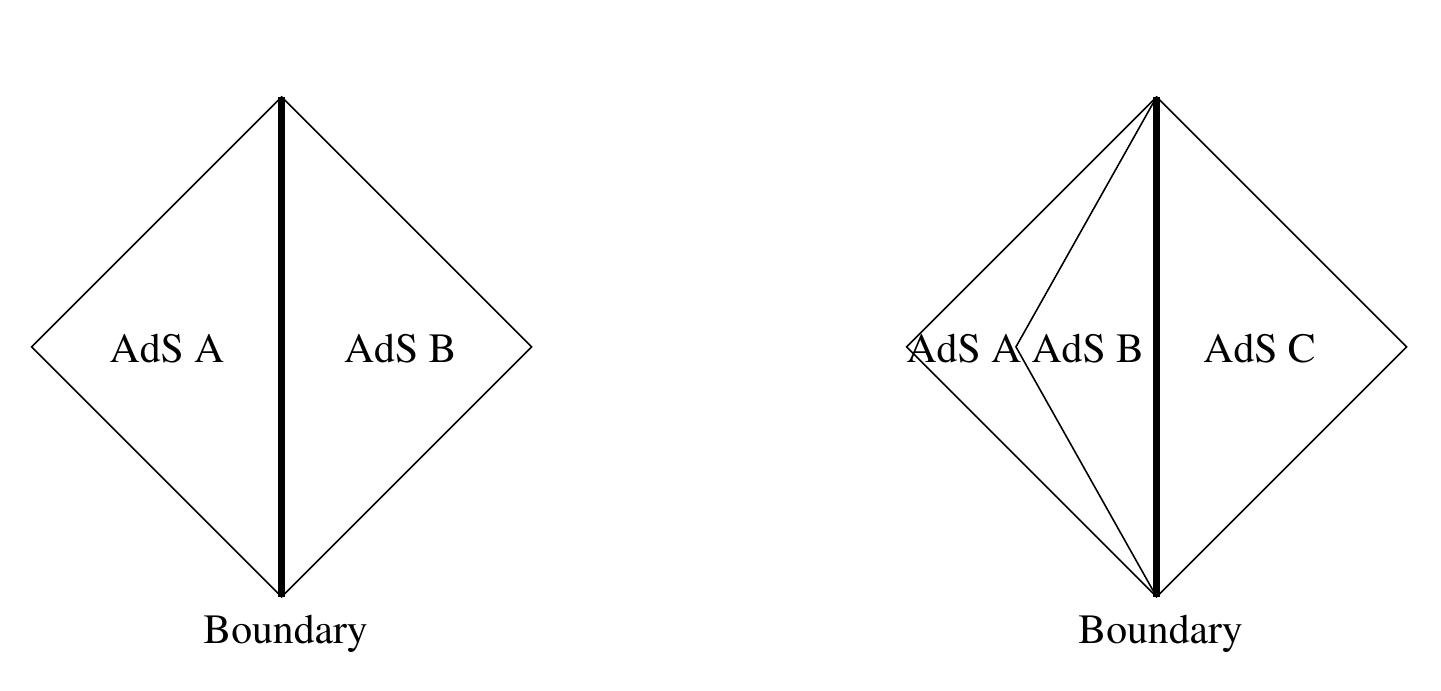}}
\caption{Schematic illustration of several CFT's with $U(1)$ global symmetries coupled to each other via simultaneously gauging all the $U(1)$'s with the same gauge field. \label{fig:junction}}
\end{figure}

Another interesting issue to consider is whether it is possible to
construct a situation where a theory in three dimensions with a two
dimensional interface is arranged so that one side of the interface is
a Witten-type theory and the other side is
some other theory in the $SL(2,{\bf Z)}$ family. One might try the
same exercise in the Marolf-Ross setup but the fact that different
theories in $SL(2,{\bf R})$ are smoothly deformable seems inconsistent
with the existence of an interface with a good UV description, unless
the interface itself has a continuous deformation parameter. These
objects are not strictly speaking `domain walls' since the
existence or absence of such objects requires more information than is intrinsic to the theory itself. One way in which
it would be possible to provide a concrete context for these objects is to require
that a bulk description exists with consistent microscopic
physics. This, presumably, can be related to the existence of a
concrete UV complete description of the interface on the boundary
side. A setup like this is very strongly reminiscent of what was
considered in \cite{Takayanagi:2011zk,Fujita:2011fp}, and it would be
of interest to explore the range of possible dynamics one can
engineer in this approach.  If the physics of the interface can be
understood sufficiently thoroughly, perhaps one might even consider
letting several boundary theories meet at a common interface and give
rise to even richer physics.

There is actually one concrete realization of an interface of this
type that one might consider. This consists of considering the ${\cal
  N}=6$ ABJM system $AdS_4 \times CP_3$ with $k$ units of RR2 flux
through $CP_1$ \cite{Aharony:2008ug}. Such a theory gives rise to a
gauge field in $AdS_4$ with gauge group $SU(4) \times U(1)$ which
contains $U(1)^4$ as an Abelian subgroup, corresponding to the global
symmetry of the boundary CFT. One can in fact think of this system as
a concrete, UV complete, realization of the CFT with global $U(1)$
symmetries for which one might consider various double trace and
contact term deformations. A particularly convenient $U(1)$ to
consider is the baryonic $U(1)_J$ current, which
\cite{Aharony:2008ug} refers to as $A_J$, which for $k^5 \gg N \gg k$
essentially corresponds to the M-theory 3-form $C_3$ integrated over
$CP_1 \in CP_3$ giving rise to a 1-form gauge field on $AdS_4$. The
IIA Chern-Simons term then essentially induces a $\theta F \wedge F$
term in the bulk $AdS_4$ where $\theta$ is the period of $B_{NSNS}$ on
$CP_1$. The bulk $F \wedge F$ term is the same thing as the boundary
Chern-Simons term. One can then think of a defect which shifts
$\theta$ as connecting two theories related by a Witten-type $T$
transformation. Since the value of $B_{NSNS}$ corresponds to the D4
Page charge \cite{Aharony:2009fc}, a D4 brane wrapped on $CP_1$ and
extended along a surface of codimension one at the boundary and the radial
coordinate will correspond precisely to a magnetic source shifting the
D4 Page charge.  Some aspect of a construction like this is
reminiscent of \cite{Fujita:2016gmu}.  It would also be interesting to
identify the defect corresponding to the $S$-transformation which
presumably is related to something resembling a Janus configuration
\cite{Gaiotto:2008sd}. All such interfaces should exhibit specific
localized dynamics which would be interesting to map out.

It would also be interesting to include the effects of gravitational
back reaction as we tune the $SL(2,{\bf R})$ and $SL(2,{\bf Z})$
boundary parameters. It is interesting to note that the $SL(2,{\bf
  Z})$ structure of Witten is somewhat sensitive to quantum issues via
incorporation of the constraint of quantization of instanton
number. It is possible that the $SL(2,{\bf R})$ of Marolf and Ross
also experiences some kind of discretization from effects which we
have not yet accounted for.

Finally, let us note that in this article, we demonstrated that there
are at least four distinct deformations, which one might call
$S_{MR}$, $T_{MR}$, $S_{W}$, and $T_W$, for the $S$ and $T$
transformations of Marolf-Ross and Witten formulations of
double trace and contact deformations. The fact that $S_{MR}$ and
$T_{MR}$ do not commute gave rise to the $SL(2,{\bf R})$ family of
theories as the maximally allowed possible deformations in the
Marolf-Ross setup, and similar consideration for the Witten setup gave
rise to the $SL(2,{\bf Z})$ structure for the set of
theories. However, it is also clear that $S_{MR}$ and $T_W$ shouldn't
commute. It could be interesting to explore the maximal set generated
by the requirement that all four generators $S_{MR}$, $T_{MR}$, $S_W$,
and $T_W$ are included. Presumably, this will be some kind of an
affine group anticipated in \cite{Casper:2017gcw}.

We hope to address some of these issues in a very near future.

\section*{Acknowledgements}

We thank  Yang Bai  and Josh Berger for discussions. This work is supported in part by the DOE grant DE-SC0017647.

\appendix

\section{Review of AdS/CFT Correspondence for bulk vector fields\label{AppA}}

The AdS/CFT correspondence for vector fields was analyzed systematically
as far back as \cite{Witten:1998qj,DHoker:1999bve,DHoker:1998bqu}.
Let us review some of the basic setup here so that the conventions and
notations are explicit. We will attempt to follow the conventions of
\cite{Casper:2017gcw} as closely as possible.

\subsection{Background geometry}

Let us consider a Euclidean Schwarzschild $AdS_{d+1}$ background with
the metric of the form
\be ds^2 = {r^2 \over R^2} (f(r) d\tau ^2 + d\vec x^2) + {R^2 \over f(r) r^2} dr^2 \ee
where
\be f(r) = 1-{r_0^d \over r^d} \ee
and $\vec x$ has $d-1$ components. We find it convenient to also introduce the radial coordinate
\be z = {R^2 \over r} \ee
so that the metric becomes
\be ds^2 = {R^2 \over z^2 } \left(f(z) d\tau ^2 + d\vec x^2+  {1 \over  f(z)} dz^2\right) \ee
with
\be f(z) = 1 - {z^d \over z_0^d}, \qquad z_0 = {R^2 \over r_0} \ . \ee
As usual,
\be T = {d \over 4 \pi R^2} r_0={d \over 4 \pi z_0}\ ,  \ee
and in these coordinates, the boundary is ${\bf R}^{d-1} \times S_1$ with the
$\tau$ coordinate being periodic
\be \tau \sim \tau + {1 \over T} \ . \ee

To this background, we add a free Maxwell field whose action reads
\be S = -{1 \over 4 e^2} \int d^{d+1}x  \sqrt{g} g^{\sigma\mu} F_{\mu \nu}  g^{\nu \lambda}F_{\lambda \sigma} \label{action} \ee
where the sign is chosen so that $S$ is positive definite in Euclidean
signature.  Here, $e^2$ has mass dimension $(3-d)$
and $A_\mu$ has dimension one. In order to state the usual AdS/CFT interpretation, we need to deal with the issue of gauge redundancy in the bulk. A standard practice is to impose the radial gauge
\be A_z = 0 \ee
and further impose the residual condition (adopting a convention that Greek indices e.g.\ $\mu$ take values $0 \ldots d$ whereas Roman indices e.g.\ $i$ take values $0 \ldots (d-1)$ excluding the $z$ coordinate)
\be \left. \partial_i A_i \right|_{z=0} = 0 \label{residual} \ee
at the boundary $z=0$ in order to fix the gauge completely.  Then, we
can interpret the remaining $A_i$ as corresponding to an operator
$O_i(x)$. We say that the $U(1)$ gauge invariance of $A_\mu$ in the
bulk is manifested as the $U(1)$ global symmetry of the field theory
dual, and that the $O_i(x)$ is the corresponding conserved current
operator.

In a realistic AdS/CFT construction which follows from string theory,
$A_\mu$ couples to other fields, especially the gravitons. In this
article, we will only discuss the toy model where the $A_\mu$ field in
the bulk is free. For the full dynamics, for concrete gauge gravity duals,
the interactions definitely matter, but they can be analyzed
separately. Our aim here is to classify the double trace and related
boundary conditions. We will also restrict attention to the two-point functions which are not as sensitive to these non-linear issues. 

\subsection{Equations of motion}

In order to carry on with the computation of the correlation function of $O_i(x)$'s, it is useful to solve the bulk wave equation as a function of the radial variable $z$ in the momentum space basis for the boundary coordinates $(t,\vec x)$. Without any loss of generality, we can orient the momentum to be of the form
\be \vec p = (\omega, p, \vec 0) \ee
using the residual $SO(d-1)$ symmetry of ${\bf R}^{d-1} \times S_1$.

It is convenient to express the action and the equation of motion in terms of momentum modes with normalization and dimensions explicitly specified. Let us define
\be \tilde A_i(z) = {T \over V_{d-1}} \int d^d x \ e^{i p x} A_i(x,z) \ee
so that
\be [\tilde A_i] = [A_i] = 1 \ . \ee
The $p$ dependence of $\tilde A_i(z)$ will be suppressed to prevent clutter.

In these conventions, the equations of motion inferred from the variation of (\ref{action}) read
\be \omega \tilde A_0'(z)+p f(z) \tilde A_1'(z) \label{eqAz} \ee
\be
\tilde A_0''(z)-\frac{(d-3) \tilde A_0'(z)}{z}-\frac{p^2
   \tilde A_0(z)}{f(z)}+\frac{\omega p \tilde A_1(z)}{f(z)} \label{eqA0}
\ee
\be
\tilde A_1''(z)+\tilde A_1'(z)
   \left(\frac{3-d}{z}+\frac{f'(z)}{f(z)}\right)-\frac{\omega^2
     \tilde A_1(z)}{f(z)^2} +
   \frac{\omega p \tilde A_0(z)}{f(z)^2} \label{eqA1}
\ee
\be
\tilde A_\perp''(z)+\tilde A_\perp'(z)
   \left(\frac{3-d}{z}+\frac{f'(z)}{f(z)}\right)-\frac{\tilde A_\perp(z)
     \left(p^2 f(z)+\omega^2\right)}{f(z)^2} \ . \label{eqA2} 
\ee
Not all of these equations are independent. We see that the $d-2$ transverse components $\tilde A_\perp(z)$ are decoupled. The independence of the remaining three equations (\ref{eqAz}), (\ref{eqA0}), and (\ref{eqA1}) can be made more explicit by changing variables to gauge and longitudinal components
\be \tilde A_g(z) = {\omega \tilde A_0 + p \tilde A_1 \over \sqrt{\omega^2+p^2}}, \qquad \tilde A_l(z) = {p \tilde A_0 - \omega \tilde A_1 \over \sqrt{\omega^2 + p^2}} \ . \ee
Then, we see that  (\ref{eqAz}) --(\ref{eqA1}) is equivalent to
\be 
\tilde A_l''(z)-{d -3 \over z} \tilde A_l'(z) +\frac{\omega^2  f'(z)}{f(z) \left(p^2 f(z) +\omega
   ^2\right)}\tilde A_l'(z) -\frac{ \left(p^2 f(z)+\omega ^2\right)}{f(z)^2} \tilde A_l(z)\label{Al} 
\ee
\be \tilde A_g'(z)=\frac{\omega  p (f(z)-1) \tilde A_l'(z)}{p^2 f(z)+\omega ^2} \label{gaugecomponent} 
\ee
which decouples the longitudinal mode into a second order differential equation, and $\tilde A_g'(z)$ becomes a first order equation. The initial condition for $\tilde A_g'(z)$ is fixed by the residual gauge condition (\ref{residual}) at $z=0$, so once $\tilde A_l(z)$ is determined, $\tilde A_g(z)$ is determined uniquely in this gauge. It is from the analysis of $\tilde A_g(z)$ where one infers that
\be \langle \partial_i O_i(x) \rangle = 0 \ee
in the usual AdS/CFT dictionary, confirming that the operator $O_i(x)$ is conserved. 

\subsection{Correlation Functions}

In order to proceed with the conventional analysis of the correlation function, we first observe that near the AdS boundary at $z=0$, we expect\footnote{In a subtle way, I claim this is consistent with the convention of (1.5) of \cite{Casper:2017gcw}.}
\be \tilde A_i(\tau,\vec x) =  \tilde a_i + \tilde O_i z^{d-2} \ee
where
\be \tilde O_i = {T \over V_{d+1}} \int d^d x\ e^{i p x} O_i(\tau,x) \ . \ee

In the standard AdS/CFT context, we can associate the operator with the variation with respect to the source
\be \tilde O_i \leftrightarrow  {1 \over {\cal N}(d-2)} {\delta \over  \delta \tilde a_i} \label{OddA} \ee
where the factor of ${\cal N}$ also shown in (\ref{calN}) can be
traced in a manner analogous to what we considered for the scalar. At
this point, dependence on $V_{d+1}$ and $R$ has been scaled entirely
into ${\cal N}$. We can therefore treat all dimensionful objects as
being dimensionless in units where $z_0$ is set to one, or restore
dimensions in appropriate powers of $z_0$, as needed.

We also need to impose the regularity of the fluctuating fields near the horizon at $z=z_0$. Solving (\ref{eqA2})  and (\ref{Al}) near $z=z_0$ reveals two linearly independent behaviors
\be \tilde A_i \sim (z_0-z)^{\pm {z_0 \omega \over d}} h(z_0-z) \ee
where $h(z_0-z)$ is analytic in $(z_0-z)$. For regularity, then, we chose the solution which decays at $z=z_0$. On Euclidean ${\bf R}^{d-1} \times S_1$, $\omega$ is quantized as
\be \omega =  {d \over 2 z_0}  n, \qquad n \in {\bf Z} \ . \ee
This makes
\be \tilde A_i \sim (z_0-z)^{n/2} h(z_0-z) \ee
which corresponds precisely to the discretization which makes $\tilde A_i$ smooth locally near the horizon. When analytically continued to Lorentzian signature in the $S_1$ direction, the condition for picking $\pm$ corresponds to picking in-falling versus out-going boundary condition at the horizon.

With the regularity condition imposed at the horizon, the solutions to the second order differential equations  (\ref{eqA2}) and (\ref{Al}) are uniquely specified upon determining the $\tilde a_i$'s at $z=0$. This means the $\tilde O_i$ are determined in terms of the $\tilde a_i$'s. Furthermore, to the extent that we are working in the probe approximation where the equations of motion are linear, the dependence of $\tilde O_i$ on $\tilde a_i$ is linear. The two-point correlation function is determined by computing
\be \langle \tilde O_i(p) \tilde O_j(-p) \rangle  = {1 \over {\cal N} (d-2)} {\delta \over \delta \tilde a_i(-p)} \langle \tilde O_j(-p) \rangle \ . 
\ee
One can further use the space-time symmetries to constrain the form of this correlation function. Had we been working on ${\bf R}^d$, we would expect the normalized two-point function
\be \Delta_{ij} = {\cal N} (d-2) \langle \tilde O_i(p) \tilde O_j(-p) \rangle  = F(p^2) P_{ij} \label{Teq0} \ee
to depend on a single momentum dependent function $F(p^2)$ where 
the projection operator is
\be P_{ij} = \left(\delta_{ij} - {p_i p_j \over p^2} \right) \ . \ee
When working instead on ${\bf R}^{d-1} \times S_1$, we expect  $\Delta_{ij}$ to break up into gauge, longitudinal, and transverse components. One can parameterize the non-trivial components of $\Delta_{ij}$ in a $2 \times 2$ matrix of the form
\be \left[\begin{array}{cc} \Delta_{ll} & \Delta_{l \perp} \\ \Delta_{\perp l} & \Delta_{\perp \perp} \end{array} \right] = \left[ \begin{array}{cc} F & pJ + pH \\ pJ-pH & G \end{array}\right] \label{2x2} \ee
and encode all the non-trivial components of the two-point function without loss of generality. The factor of $p$ in front of $J$ and $H$, corresponding to the symmetric and anti-symmetric off-diagonal components, respectively,  is for future convenience. So, we find that the symmetry of ${\bf R}^{d-1} \times S_1$ allows up to four independent components of $\Delta_{ij}$, corresponding to four components of a $2 \times 2$ matrix. 

For the pure Maxwell system whose equations of motion are (\ref{eqA2}) and (\ref{Al}), the longitudinal and the transverse components are decoupled, so $J=H=0$. The only non-trivial components are $F$ and $G$. (In the $T\rightarrow 0$ limit, $F$ and $G$ become identical and reduces to (\ref{Teq0}).)

\begin{figure}
  \centerline{  \includegraphics[width=3in]{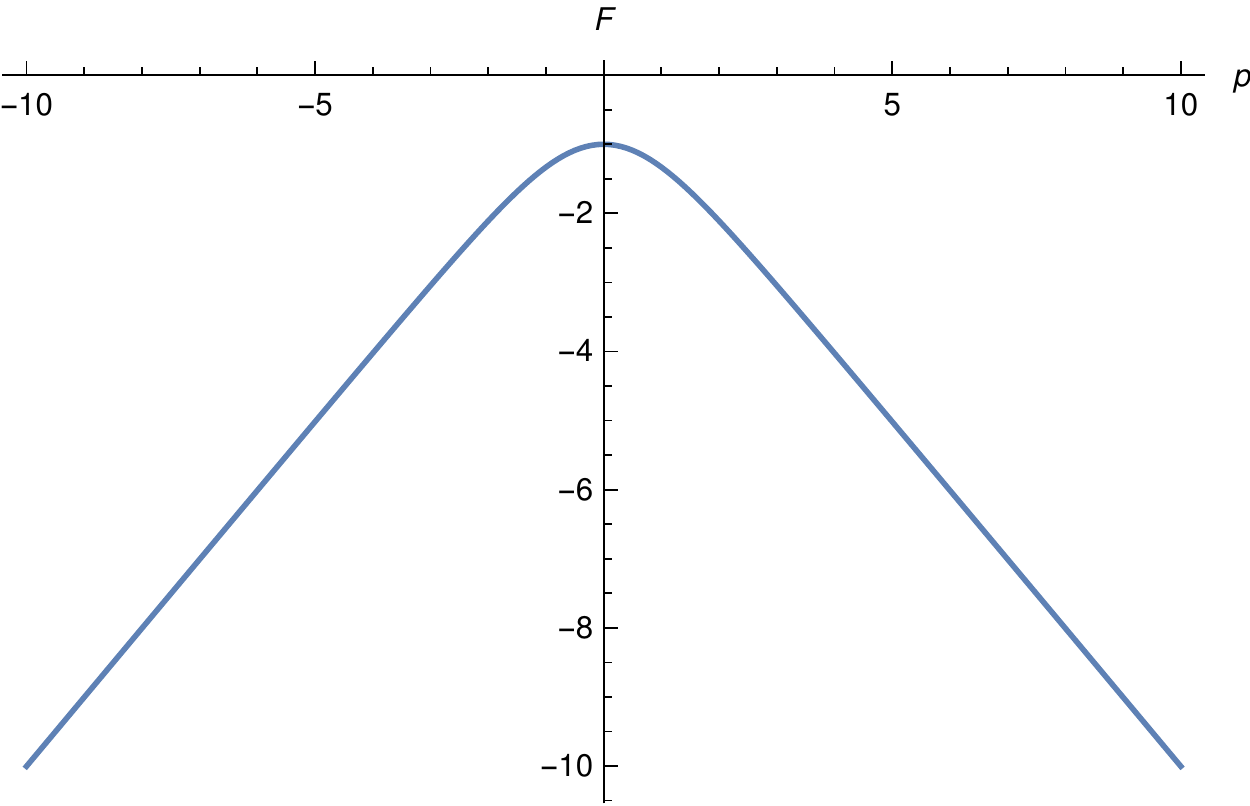}  \includegraphics[width=3in]{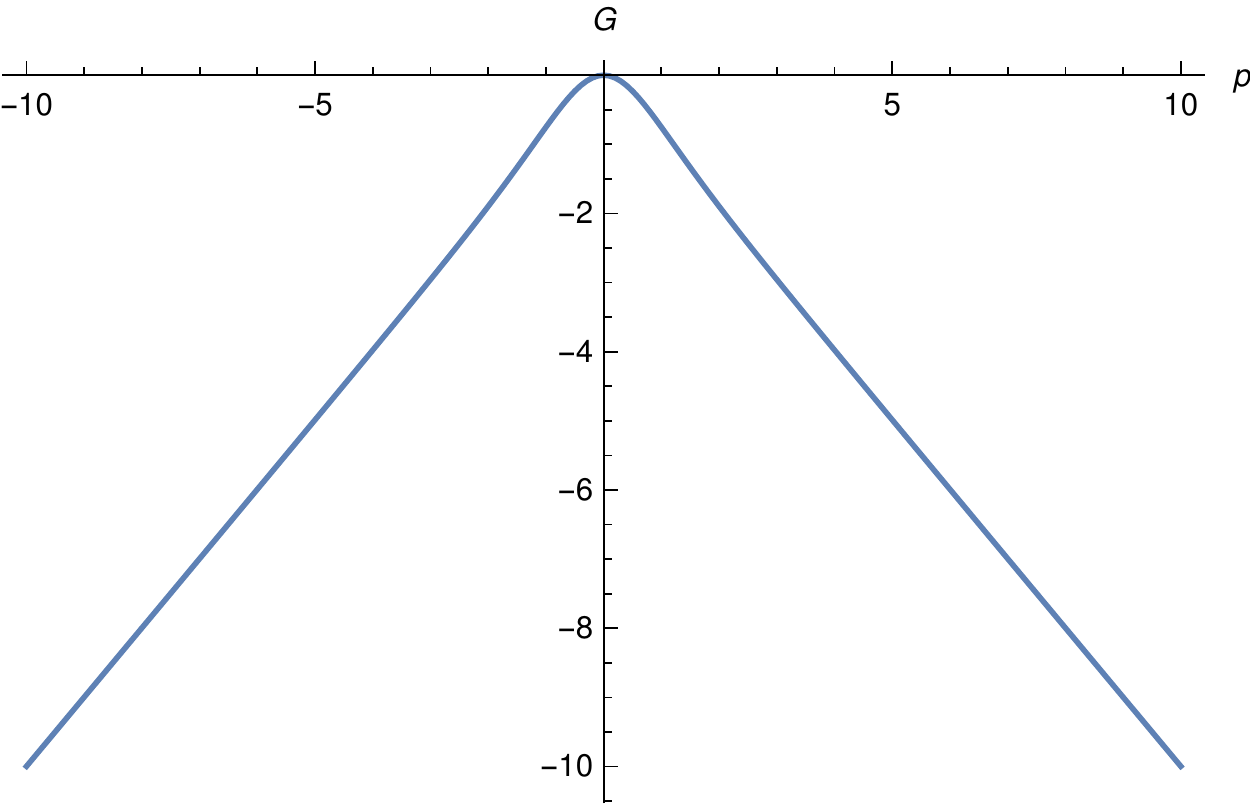}}
  \caption{Correlation function $F=\Delta_{ll}(p^2,\omega)$ and
    $G=\Delta_{\perp \perp}(p^2,\omega)$ plotted as a function of $p$
    with $\omega=0$ for the Maxwell field in AdS-Schwarzschild
    background with standard Dirichlet boundary condition in units
    where $z_0=1$. \label{figA}}
\end{figure}

All that remains is to determine $F$ and $G$ by solving (\ref{eqA2}) and (\ref{Al}). Unfortunately, these equations are not analytically solvable, but can easily be solved numerically and analyzed in the small $p^2$ limit. The result of this analysis, for the case where we set $d=3$ and $\omega=0$, is illustrated in figure \ref{figA}.

Note that both $F(p^2)$ and $G(p^2)$ are negative.  These plots should
be viewed as the analogue of figure 3 of \cite{Casper:2017gcw} except
that here, we are plotting the correlation function for the Dirichlet,
rather than the Neumann theory. In fact, $F(p^2)$ behaves
qualitatively the same way as the scalar two-point function. This is
not unexpected, in that $A_0$ behaves as a Kaluza-Klein scalar on ${\bf
  R}^{d-1} \times S_1$. The transverse correlation function encoded by
$G(p^2)$ is different from $F(p^2)$, in that $G(p^2)=0$ for $p=0$.

We can also compute the small $p^2$ asymptotic behavior
\be F_D(p^2=0) = - 1 + {\cal O}(p^2), \qquad G_D(p^2=0) = -  p^2 + {\cal O}(p^4) \ . \ee
Explicit computations reveal that the coefficient of the constant term
in $F_D$ and the $p^2$ term in $G_D$ are precisely $-1$.  As was noted
shortly below (\ref{OddA}), $F$ and $G$ can be treated as being
dimensionless by working in units where $z_0=1$.

\bibliography{spin1}\bibliographystyle{utphys}

\end{document}